\documentclass[prb,twocolumn,amsmath,amsymb,showpacs]{revtex4}

\usepackage{graphicx}
\usepackage{dcolumn}
\usepackage{bm}

\begin{document}

\title{In search of antiferromagnetic interlayer coupling in diluted magnetic thin films with RKKY interaction}

\author{Karol Sza{\l}owski}
\email{kszalowski@uni.lodz.pl}
\author{Tadeusz Balcerzak}%
\affiliation{%
Department of Solid State Physics, University of \L\'{o}d\'{z},\\
ulica Pomorska 149/153, 90-236 \L\'{o}d\'{z}, Poland
}%

\date{\today}

\begin{abstract}
We study a model thin film containing diluted bilayer structure with the RKKY long-range interaction. The magnetic subsystem is composed of two magnetically doped layers, separated by an undoped nonmagnetic spacer and placed inside a wider film modelled by a quantum well of infinite depth. We focus our study on the range of parameters for which the antiferromagnetic coupling between the magnetic layers can be expected. The critical temperatures for such system are found and their dependence on magnetic layer thickness and charge carriers concentration is discussed. The magnetization distribution within each magnetic layer is calculated as a function of layer thickness. The external field required to switch the mutual orientation of layer magnetizations from antiferromagnetic to ferromagnetic state is also discussed.
\end{abstract}

\maketitle

\section{Introduction}

The phenomenon of mutual coupling of magnetic layers separated by a nonmagnetic medium has been of great interest since its discovery in Fe/Cr structure,\cite{Grunberg} both due to its immense application potential and the highly nontrivial physical background.\cite{grunbergJMMM,BrunoC,Barnas1,Sapozhnikov,dugaev,Edwards2,Staniucha}. In parallel to studies of metallic systems, the progress in the field of thin films made of diluted magnetic semiconductors (DMS), such as the most representative (Ga,Mn)As, offers some new possibilities to study the diluted magnetic systems with confined geometry. Such materials are promising from the future spintronics point of view,~\cite{Fabian} since they can assure integrability of novel spin-current devices with conventional structures of semiconductor-based electronics.\cite{Fert} In DMS an indirect coupling between localized spins mediated by charge carriers is of paramount importance for the possibility of magnetic ordering. The Ruderman-Kittel-Kasuya-Yosida (RKKY) interaction mechanism was suggested to describe the properties of quantum-well based DMS systems a decade ago.~\cite{DietlPRB}

The problem of interlayer coupling in multilayers and superlattices of DMS has been subject to various theoretical studies. Giddings \emph{et al.}\cite{giddingspssc} performed some calculations within a single parabolic band approximation for $\mathbf{k}\cdot \mathbf{p}$ method and local density approximation to study the possibility of the appearence of antiferromagnetic (AF) interlayer coupling in the superlattices of (Ga,Mn)As. They concluded that the minimization of spacer thickness accompanied with relatively short superlattice period is advantageous for a strong coupling. This problem has also been investigated by Sankowski and Kacman,~\cite{Sankowskicoupling}  who found in their tight binding-based study a relative insensitivity of the coupling energy to the magnetic layer thickness. A noticeable range of parameters favourizing the AF interlayer interaction has been identified. The self-consistent mean-field calculations for multilayers were also performed by Jungwirth \emph{et al.}\cite{Jungwirthcoupling}

Numerous experimental works evidenced the fact that two magnetically doped layers can be indirectly coupled for some interlayer distances. For example, such a coupling has been observed in (Ga,Mn)As/GaAs trilayers by means of magnetotransport and magnetization process studies by Akiba \emph{et al.}\cite{akibacoupling}, where the quantitative analysis was performed yielding the ferromagnetic (F) interlayer couplings close to 0.1 mJ/m$^{2}$ for the nonmagnetic spacer up to 10 ML thick. The superlattice structures of the same DMS have been investigated by Mathieu \emph{et al.}\cite{mathieucoupling}, where the critical temperature was found to oscillate with the spacer thickness (up to 9 ML) and thus the presence of interlayer coupling was deduced. Short-period superlattices formed the basis of neutron reflectivity studies of Szuszkiewicz \emph{et al.}\cite{szuszkiewiczcoupling1}, who confirmed the ferromagnetic-only interlayer coupling (mainly for 6 ML spacer). In the work of Kirby \emph{et al}.\cite{kirbycoupling} neutron reflectometry and magnetization curve studies of trilayer structures were performed, in which the magnetization of top and bottom layer has been resolved. This indicated distinct ferromagnetic coupling through a spacer of large thickness (about 20 ML). It was also excluded that the spacer has been plagued by diffusion of the magnetic impurities leading to direct coupling mechanism. On the other hand, the results of Ge \emph{et al.}\cite{AFcouplingjunctions} show the difference of critical temperatures for top and bottom layer in a trilayer system, a difference which had a tendency to vanish when the nonmagnetic spacer thickness was reduced (and a single common transition temperature has been observed for the spacer of about 10 ML thick). Let us note that in that paper it has been stated that the sign of coupling energy might be spatially inhomogenous - either F or AF for various areas on the surface of the sample, which was deduced from the observation of planar Hall effect. However, it was not until last year that the AF coupling between layers in (Ga,Mn)As/GaAs:Be was clearly observed by means of DC magnetometry as well as neutron reflectivity studies performed by Chung \emph{et al}.\cite{chungAF} The critical temperature of the system was about 50 K. All the discussed interlayer coupling signatures have been assigned to the carrier-mediated interaction mechanism. 

The above results encourage for further search for the possibility of obtaining AF interlayer carrier-mediated coupling. Therefore, it seems purposeful to devote a theoretical study to the trilayer structure, which can be formed by two magnetic layers immersed in the thin film and separated by a nonmagnetic spacer, with the aim to discover some unique properties of RKKY interaction in the ultrathin film. From the practical point of view it is interesting to find a range of carrier concentrations and geometric characteristics with AF coupling. In particular, the influence of the magnetic layer thickness on the coupling energy and the critical temperature of the system poses an important question. In order to have a deeper insight, also the spatial distribution of magnetization within the magnetic layers appears worth-while to examine.

For the above reasons, in this paper our purpose is to study a model trilayer system with the RKKY interaction. The theoretical model is based on the quantum-well approach to the thin film for which the RKKY exchange coupling has been re-derived. The numerical calculations have been performed for some realistic material constants, typical of DMS systems.

\section{Theoretical model}

\begin{figure}
\includegraphics[scale=1.0]{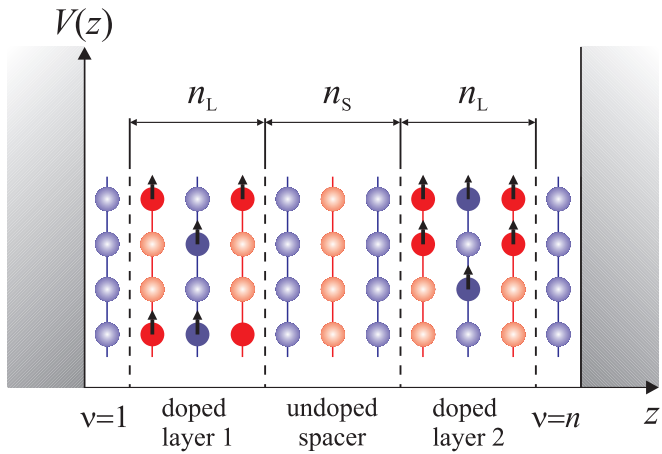}
\caption{Schematic cross-section of an ultrathin film containing two magnetically doped layers. In this figure the total thickness of the film amounts $n$=11, whereas the thickness of each magnetic layer is $n_{L}$=3 and the thickness of non-magnetic spacer is $n_{S}$=3.}
\label{fig:fig1}
\end{figure}
We consider a model thin film consisting of $n$ monolayers, where some of the planes can be doped with magnetic impurity ions. The doped planes form two magnetic layers,  separated by a non-magnetic spacer (Fig.1). Each magnetic plane consists of $n_{L}$ doped atomic planes, while the undoped spacer has the thickness of $n_{S}$ monolayers. As a consequence, the inequality $2n_{L}+n_{S}\leq n$ must hold. The film possesses fcc structure and its surface orientation is (001). The localized spins situated in the lattice site $\left(\mathbf{r}_i,z_{\nu}\right)$ can be shortly denoted by $\mathbf{S}_{i,\nu}$, where $\nu$ is one of the atomic planes whose position is given by $z_{\nu}=(\nu - 1/2)d$, and $\mathbf{r}_i$ is the vector in the plane. $d$ is the thickness of a single monolayer, and the total thickness of the film is $D=nd$. The choice of structure is motivated by the fact that the ultrathin films of (Ga,Mn)As are most often grown with this surface orientation, on (001)GaAs substrate. Our film constitutes a quantum well of infinite depth in $z$ direction for the charge carriers. 

For the free-electron model, the perturbation Hamiltonian describing the exchange interaction between the single localized magnetic moment at $\left(\mathbf{r}_i,z_{\nu}\right)$ and the free charge carrier at $\left(\mathbf{r},z\right)$ is assumed to be in the following form:
\begin{equation}
\mathcal{H}=-A\,p\left(\mathbf{r},z\right)\,S^{z}_{i,\nu}\,s^{z}\left(\mathbf{r},z\right),
\label{eq:exchange}
\end{equation}
where $A$ is the exchange constant,
\begin{equation}
p\left(\mathbf{r},z\right)=\frac{1}{\sigma^3\left(2\pi\right)^{3/2}}\,e^{-\left(\left(r-r_i\right)^2+\left(z-z_{\nu}\right)^2\right)/2\sigma^2},
\label{eq:contactpotential}
\end{equation}
and $s^{z}\left(\mathbf{r},z\right)$ is the $z$-component of the free-electron spin in the $\left(\mathbf{r},z\right)$ point of the film.

We emphasise the fact that the selected "diffused" contact potential $p\left(\mathbf{r},z\right)$ differs from the usual Dirac delta-function, since it has a non-zero dispersion $\sigma^2$. The contact potential frequently used in derivation of RKKY indirect exchange integral is reproduced in the limit $\sigma\to 0$. Such a form of the formula (\ref{eq:contactpotential}) is further substantiated by the fact that the localized magnetic moments originate from the $d$-type electronic orbitals which possess some spatial extension. Note that such a potential has already been applied in studies of bulk DMS.\cite{Schliemann,SchliemannMC,Brey} 

The influence of "diffusion" of the contact potential on the resulting RKKY coupling integral in 1D, 2D and 3D systems has recently been studied in Ref.~\onlinecite{RKKYdiffused}.
The studies resulted in obtaining a finite value of RKKY integral for $\mathbf{r}=0$ in 2D and 3D case. The importance of this result has first been noticed for discrete 3D systems since the presence of divergency at $\mathbf{r}\to 0$ leads to some divergent self-energy corrections.\cite{Schliemann} However, it must be emphasized that in studies of ultrathin films the acceptance of a finite $\sigma$ value is necessary, even if we do not require the self-energy to be finite. The reason is that the RKKY exchange integral derived on the basis of the contact potential with $\sigma=0$ is divergent for arbitrary $z$ when $r=0$, i.e., in the direction perpendicular to the film plane, even though the distance between the interacting spins is nonzero. This fact precludes the further studies of the total energy of the discrete system if $\sigma=0$.

By applying the second-order perturbation calculus we obtain the RKKY exchange integral between the localized spins separated by the distance $\Delta r=|\mathbf{r}_i - \mathbf{r}_j|$ in the plane of the film, and $\Delta z=|z_{\nu} - z_{\mu}|$ in the direction perpendicular to the film surface. The integral has the form of:  
\begin{eqnarray}
J\left(\Delta r,\Delta z\right)&=&C\left(\frac{a}{d}\right)^4\frac{\pi}{4n^2}\int_{0}^{+\infty}\!\!\! dy{\left[yJ_{0}\left(y\frac{\Delta r}{d}\right)e^{-\left(\frac{\sigma}{d}\right)^2y^2}\right.}\nonumber\\
&&\times\left.\sum_{l=-\infty}^{+\infty}{\frac{\chi_{l,y}}{\chi_{0}}\cos\left(\frac{2\pi}{n}\frac{\Delta z}{d}l\right)\phi^{2}\left(l\right)}\right].
\label{eq:rkky_tf}
\end{eqnarray}
In the above formula, $\phi\left(l\right)$ is the Fourier transform of the $z$-dependent part in the interaction potential $p\left(\mathbf{r},z\right)$, and it can be presented as $\phi\left(l\right)=\exp\left(i\frac{2\pi}{D}l\sigma \right)$ for $\sigma/d\ll 1$.\cite{Jaszowiec} The symbol $\chi_{l,y}$ denotes the paramagnetic susceptibility of the electrons in the ultrathin film, which has been derived elsewhere,\cite{BalcerzakTF2,BalcerzakJMMM} while $\chi_{0}$ is the Pauli susceptibility of 2D system. The energy constant $C$ in Eq.(3) is given by $C=2mA^2/\pi h^2a^4$, where $a$ is the lattice constant.

The Hamiltonian for the system under consideration is assumed to be of Ising-type with the long-range exchange interaction $J\left(\Delta r,\Delta z\right)$: 
\begin{eqnarray}
\mathcal{H}&=&-\sum_{\left\langle i,\mu;j,\nu\right\rangle}^{}{\! J\left(\Delta r,\Delta z\right)\,\xi_{i,\mu}\xi_{j,\nu}S^{z}_{i,\mu}S^{z}_{j,\nu}}\nonumber\\
&+&g_{\rm eff}\mu_{\rm B}B\sum_{i,\mu}^{}{\xi_{i,\mu}S^{z}_{i,\mu}}.
\label{eq:hamiltonian}
\end{eqnarray}
The above Hamiltonian includes the Zeeman term for the system embedded in an uniform external field $B$. The localized spins reveal the effective gyromagnetic factor $g_{\rm eff}=g_S+g_e (\pi Am\tau_{\rm F}/h^2D)$,\cite{BalcerzakRKKY} where $g_S$ and $g_e$ are the gyromagnetic factors of localized spins without RKKY interaction and the free-electrons, respectively. $\tau_{\rm F}$ is the number of subbands, splitted due to the discretization of Fermi surface in thin film.\cite{BalcerzakTF1}

In order to describe a site-dilution in the system we make use of the Edwards operators $\xi_{i,\mu}$,\cite{edwards} which possess two eigenvalues: 1 when the site $(i,\mu)$ is occupied by a magnetic impurity, or 0 otherwise. 
In this paper, for the simplicity, we will neglect the correlations of $\xi_{i,\mu}$ operators, i.e., the Warren-Cowley short-range-order parameter~\cite{RKKYcorrelations} is assumed here to be 0. It means that the occupation of lattice sites by the impurity ions in doped planes is completely random.
As a consequence, the configurational averages read $\left\langle\xi_{i,\mu}\right\rangle_{r}=x$ and $\left\langle\xi_{i,\mu}\xi_{j,\nu}\right\rangle_{r}=x^2$, if the sites $(i,\mu)$ and $(j,\mu)$ belong to the magnetically doped atomic planes and vanish otherwise, whereas $x$ is the magnetic dopant concentration.  In further considerations it will be assumed that the indices $\mu$ and $\nu$ run only over the magnetically doped atomic planes.

By taking thermodynamical and configurational average of the Hamiltonian (\ref{eq:hamiltonian}) for the temperature $T\to 0$, we obtain the ground-state enthalpy $H=<\mathcal{H}>$ of the magnetic system under consideration. The enthalpy can be conveniently written as a sum of contributions originating from the energy of spin interactions within a single magnetic layer, the energy of interactions between two layers and the interaction with an external field. For the temperature $T\to 0$ the magnetization in each atomic plane $\nu$ constituting a specified magnetic layer is the same, for instance, $\left\langle S^{z}_{i,\mu}\right\rangle=m_{1}=\pm S$ for the first layer and $\left\langle S^{z}_{i,\mu}\right\rangle= m_{2}=\pm S$ for the second one, i.e., both layers are fully polarized (ferro- or antiferromagnetically each to the other). As a consequence, the total enthalpy per one site in a doped atomic plane reads:
\begin{equation}
\frac{H}{N}=\frac{E_{\mathrm{intra}}}{N}+\frac{E_{\mathrm{inter}}}{N}-xn_{L}g_{\rm eff}\mu_{\rm B}B\left(m_{1}+m_{2}\right).
\label{eq:entalphy}
\end{equation}
Since each layer is ferromagnetically ordered inside, the energies of intralayer and interlayer interactions can be written in the form of:
\begin{equation}
\frac{E_{\mathrm{intra}}}{N}=-x^2S^2\,\mathcal{E}_{\mathrm{intra}}
\label{eq:intralayer}
\end{equation}
and
\begin{equation}
\frac{E_{\mathrm{inter}}}{N}=-x^2m_{1}m_{2}\,\mathcal{E}_{\mathrm{inter}}.
\label{eq:interlayer}
\end{equation}
In the above equations we denoted the appropriate lattice sums by:
\begin{equation}
\mathcal{E}_{\mathrm{intra}}=n_{L}\mathcal{E}\left(0\right) + \sum_{\nu=1}^{n_{L}-1}{\left(n_{L}-\nu \right)\mathcal{E}\left(\nu \right)}
\label{eq:intralayer2}
\end{equation}
and
\begin{eqnarray}
&&\mathcal{E}_{\mathrm{inter}}=n_L \mathcal{E}\left(n_{L}+n_{S}\right)\nonumber \\
&\!\!\!\!\!\!\!\!\!\!\!\!+&\!\!\!\!\!\!\!\sum_{\nu=1}^{n_{L}-1}{\left(n_{L}-\nu\right) \left[ \mathcal{E}\left(n_{L}+n_{S}+\nu\right)+\mathcal{E}\left(n_{L}+n_{S}-\nu\right)\right]},
\label{eq:interlayer2}
\end{eqnarray}
where
\begin{equation}
\mathcal{E}\left(\omega\right)=\sum_{k}^{}{z^{\mathcal{P}\left(\omega\right)}_{k}J\left(r^{\mathcal{P}\left(\omega\right)}_k,\omega d\right)}. 
\label{eq:sum}
\end{equation}

The logic index $\mathcal{P}\left(\omega\right)$ stands for the parity of $\omega$, which refers to the fact that in (001)-oriented fcc-based thin film, the subsequent atomic planes are mutually shifted parallelly to the surface. This requires two sets of co-ordination zones radii $r_k$, and co-ordination numbers $z_k$, to be used: one if both spins are situated in the same plane (or in the planes seperated by an even number of interplanar distances), $\mathcal{P}(\omega)={\rm "even"}$, and the other for the situation when both spins lie in the planes seperated by an odd number of interplanar distances, $\mathcal{P}(\omega)={\rm "odd"}$. Both sets of numbers have to be determined numerically for the assumed crystallographic structure. $\mathcal{P}\left(0\right)$ requires the point $\left(\mathbf{r}=0,z=0\right)$ to be excluded from summation.

In the system considered, the area per one site in the (001) atomic plane is $S=a^2/2$ so that the interlayer coupling energy per unit area of the film is $2x^2S^2\mathcal{E}_{\mathrm{inter}}/a^2$.

Let us note that the interlayer coupling energy is often defined as the difference in energies of the system for ferromagnetic ($E^{\uparrow\uparrow}$) and antiferromagnetic ($E^{\uparrow\downarrow}$) alignment of magnetic moments in these layers. In our notation such a difference is $\left(E^{\uparrow\uparrow}-E^{\uparrow\downarrow}\right)/N=-2x^2S^2\mathcal{E}_{\mathrm{inter}}$.

When we consider the magnetic monoatomic layers with the thickness $n_{L}=1$, the expressions (\ref{eq:intralayer2}) and (\ref{eq:interlayer2}) reduce only to the first terms, without summation over $\nu$.\cite{Jaszowiec} On this basis, as a special case, we can study the interplanar coupling energy for two magnetically doped planes as a function of their separation for different total thicknesses of the film.

In general, for non-zero temperature, the magnetizations $m_{\mu}$ per one magnetic impurity in each magnetically doped plane are not equal for all $\mu$, so that some spatial magnetization profile is present in the magnetic layers. In the molecular field approximation (MFA), these magnetizations can be obtained as the solutions to the set of $2n_L$ coupled self-consistent equations
\begin{equation}
m_{\mu}=S\mathcal{B}_{S}\left(S\frac{\Lambda_{\mu}}{k_BT}\right).
\end{equation}
It should be kept in mind that the index $\mu$ here runs only over the magnetically doped monolayers.
$\mathcal{B}_{S}\left(x\right)$ is the Brillouin function for spin $S$ and the molecular field acting on a given spin in the layer $\mu$ is
\begin{equation}
\Lambda_{\mu}=x\sum_{\nu}^{}{\mathcal{E}\left(\left|\mu-\nu\right|\right)m_\nu}.
\end{equation}
The summation over $\nu$ in Eq.(12) is restricted to the magnetically doped atomic planes.

From linearization of the above set of equations for magnetization, in vicinity of the critical point, we obtain the critical temperature of a second-order phase transition. It is given by the largest real root of the equation:
\begin{equation}
\mathrm{det}\,\mathcal{J}=0,
\label{eq:tc}
\end{equation}
where the matrix $\mathcal{J}$ is of size $2n_{L}\times 2n_{L}$, and its elements are
\begin{equation}
\mathcal{J}_{\mu\nu}=\delta_{\mu\nu}-\frac{S\left(S+1\right)}{3k_{B}T_{\rm c}}\mathcal{E}\left(\left|\mu-\nu\right|\right).
\label{eq:tcmatrix}
\end{equation}

Due to the long-range summation occurring in $\mathcal{E}\left(\left|\mu-\nu\right|\right)$ (as seen from Eq.(10)), the critical temperature can be calculated from the equation (\ref{eq:tc}) only numerically. The formula given above can be approximated if we assume an uniform magnetization distributions within each of the magnetic layers, i.e., when we substitute $m_1$ for the average magnetization in each atomic plane inside the magnetic layer 1 and $m_{2}$ for the layer 2, respectively. Then the critical temperature can be calculated from the approximate formula
\begin{equation}
k_{\rm B}T_{\rm c}=\frac{S\left(S+1\right)}{3}\frac{2\mathcal{E}_{\mathrm{intra}}-n_{L}\mathcal{E}\left(0\right)\pm\mathcal{E}_{\mathrm{inter}}}{n_{L}},
\label{eq:tcapproximate}
\end{equation}
where the "$+$" sign corresponds to the Curie temperature and is valid for ferromagnetic interlayer coupling ($E_{\mathrm{inter}}<0$), while the "$-$" sign is for the N\'{e}el temperature and is valid when the magnetic layers are coupled antiferromagnetically ($E_{\mathrm{inter}}>0)$. The highest accuracy of the uniform approximation (15) is for small magnetic layer thicknesses $n_{L}$.  

When AF coupling exists between two magnetic layers, the mutual orientation of their magnetizations can be switched from antiparallel (antiferromagnetic) to parallel (ferromagnetic) one by applying an external magnetic field along the direction of their magnetizations. This kind of field-induced phase transition can take place below the N\'{e}el temperature. The critical field $H_{\rm c}$ required to force the reversal of magnetization at $T\to 0$ can be determined by equating the total enthalpies of the system with FM and AFM orientation of magnetizations in the presence of that external field. The enthalpies are obtained from the formula (\ref{eq:entalphy}), which leads to the condition for $H_{\rm c}$:
\begin{equation}
\mu_{0}H_{\rm c}=\frac{\left|E_{\rm inter}\right|}{g_{\rm eff}\mu_{\rm B}SNxn_{L}}.
\end{equation}
The above expression can be written in the form more convenient for numerical calculations:
\begin{equation}
\mu_{0}H_{\rm c}=B_{0}\frac{x}{n_{L}}\frac{|\mathcal{E}_{\rm inter}|}{C},
\label{eq:Bcr}
\end{equation}
where
\begin{equation}
B_{0}=\frac{CS}{g_{\rm eff}\mu_{\rm B}}
\label{eq:B0}
\end{equation}
is a material-dependent constant.

\section{Numerical results and discussion}

\begin{figure}
\includegraphics[scale=0.71]{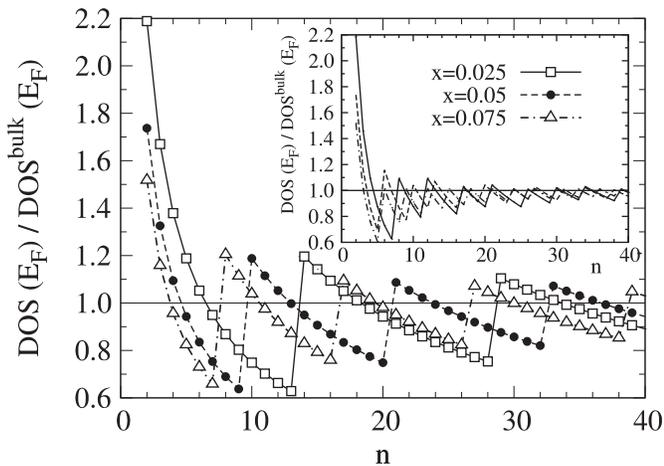}
\caption{Density of states at the Fermi level in an ultrathin film, normalized to the bulk value, as a function of number of monolayers . The charge carriers concentration is assumed $\rho=x/n$ for three representative values of $x$. In the inset, a similar plot is shown for the charge carrier concentration $\rho=x/2$.}
\label{fig:fig2}
\end{figure}

Since the existence and characteristic of RKKY interaction essentially depend on the electronic structure of the system under consideration, we make it a point of a brief discussion for the ultrathin film. Confining the charge carriers in a quantum well results in discretization of the Fermi surface, which for such a system is composed of a finite number of circular 'slices', each of them corresponding to one 2D energy subband. The detailed description of the electronic structure of an ultrathin film is presented in Ref.~\onlinecite{BalcerzakTF1}. In the Fig.~\ref{fig:fig2}, the crucial quantity for studies of electronic properties, namely the density of states (DOS) at Fermi level is plotted against the number of monolayers which make up the whole film. The method of calculating DOS for an ultrathin film has been presented in details in Ref.~\onlinecite{BalcerzakTF1}. In further studies we make use of a normalized charge carriers concentration $\rho=n_{c}d^3$. In the main plot, it is assumed that the number of charge carriers in the film is fixed and their normalized concentration equals $\rho=x/n$, thus carrier concentration decreases when the film is made thicker. This assumption corresponds to the situation in which only two out of $n$ atomic planes contain impurities which serve as charge donors, and the concentration of impurities within each doped plane is equal to $x$. Three impurity concentrations were selected, namely $x=$0.025, 0.050 and 0.075 all belonging to the physically relevant range in DMS. For the inset we chose constant $\rho=x/2$, which is valid when all the atomic planes are doped with concentration $x$. For convenience, the values of DOS were normalized to the corresponding values for the bulk case (for the appropriate value of $\rho$). 
\\
It is visible that DOS undergoes discontinuous jumps at some values of $n$. This is clear manifestation of quantum size effects (QSE) as each of the jumps occurs for the charge carrier concentration at which the next energy subband becomes occupied by the carriers, starting from a single 2D-like subband for the lowest concentrations. Although $\rho$ is inversely proportional to the film thickness, the number of occupied energy subbands increases while increasing $n$. The distance between subsequent jumps becomes shorter by increasing the concentration $x$ of charge carrier donors. Between the jumps, normalized DOS decreases as $n^{-2/3}$, which reflects the fact that for a thin film the total DOS is a sum of (equal) contributions from each occupied 2D-like subband. This 2D-like DOS per one charge carrier is energy-independent, so that dividing by the 3D DOS per one charge carrier (proportional to $\rho^{-2/3}$) yields this behaviour. The dependence presented in the main plot in Fig.~\ref{fig:fig2} is a result of interplay between the changes caused solely by the change in quantum well width and by decrease of the carrier concentration when the film becomes thicker. In the inset, for constant carrier concentration, we observe that the magnitude of QSE tends to vanish more rapidly for thicker films. For the film thicknesses $n$ for which QSE manifest, we can expect similar discontinuous bahaviour of other physical properties which depend crucially on the DOS at the Fermi level.

\begin{figure}
\includegraphics[scale=0.71]{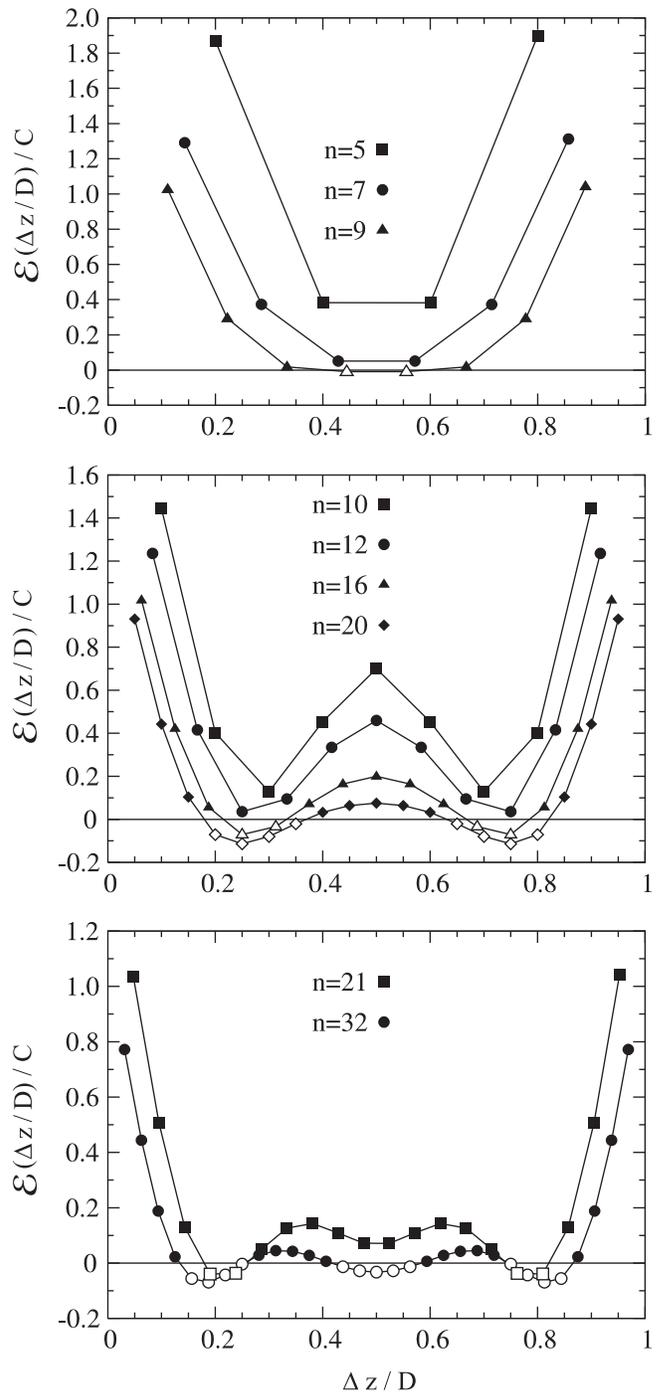}
\caption{Interplanar coupling energy for an ultrathin film containing $n$ monolayers, as a function of separation between the planes, for the charge carriers concentration $\rho=x/n$ and $x=0.05$.}
\label{fig:fig3}
\end{figure}

\begin{figure}
\includegraphics[scale=0.71]{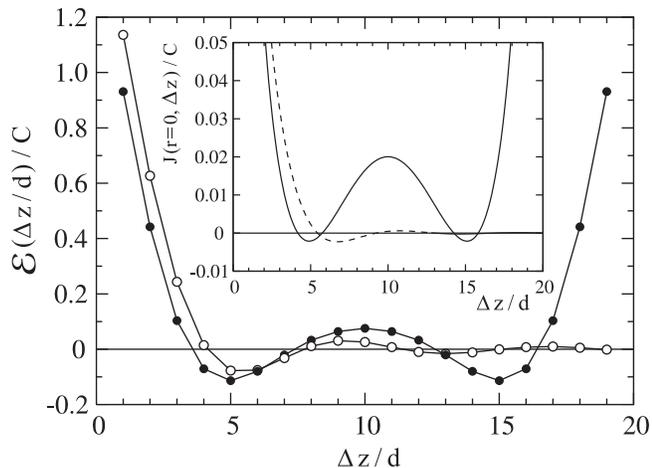}
\caption{Interplanar coupling energy in an ultrathin film for $n=20$, $x=0.05$, and the charge carriers concentration $\rho=x/n$. Full points denote the values obtained from summation of the  RKKY thin film-modified exchange integral. The empty points represent the values obtained from summation of the usual RKKY bulk formula. In the inset, the very exchange integrals $J\left(r=0,z\right)$ are plotted - the solid line is for the RKKY interaction in thin film, while the dashed line corresponds to the bulk 3D RKKY coupling.}
\label{fig:fig4}
\end{figure}

\begin{figure}
\includegraphics[scale=0.71]{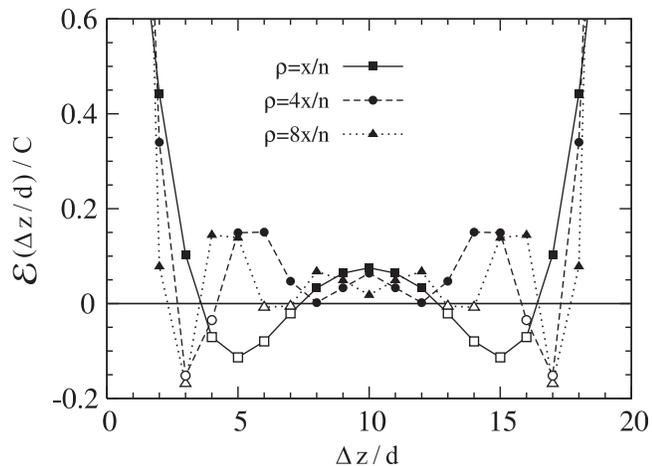}
\caption{Interplanar coupling energy for an ultrathin film characterised by $n=20$, $x=0.05$, and for three representative charge carrier concentrations.}
\label{fig:fig5}
\end{figure}

Having a special case of the formula (\ref{eq:interlayer2}) for $n_{L}=1$ (as in Ref.\onlinecite{Jaszowiec}), we can study the interplanar coupling energies $\mathcal{E}\left(\Delta z/d\right)$. In the formula (\ref{eq:sum}), the summation of exchange integrals over the in-plane coordination zones has to be performed. Some cut-off radius must be assumed, large enough to assure satisfactory convergency of $\mathcal{E}$. It should also be emphasized that calculating each value of exchange integral $J\left(r_{k},\omega d\right)$ constitutes a remarkable numerical task itself. Therefore, a noticeable computational effort is required to study the inter- and intralayer interaction energies so that we performed our calculations on the multi-CPU cluster. We performed the summation up to $k=$1000 coordination zones (which corresponds to the cut-off radius $r/d$=82.0 for even interplanar distances and $r/d$=86.8 for odd interplanar distances). In addition, we applied some averaging intended to remove the component oscillatory in cut-off radius. Such a procedure provides sufficient convergency of the resulting energies. Let us emphasize that accepting a finite spatial extension of an interaction potential (\ref{eq:exchange}) is necessary to obtain finite interplanar coupling since the formulas contain the exchange integral $J\left(0,z\right)$ which would be divergent otherwise. In further calculations we assumed $\sigma/d$=0.35, which could correspond to 1 \AA $\;$ for (Ga,Mn)As (which we assume to be a realistic estimate of the spatial extension of $d$ orbital carrying the localized spin).

It is interesting to follow the evolution of the energy dependence on interplanar distance when the film thickness $n$ increases. We make an assumption that the charge carrier concentration is $\rho=x/n$. In Fig.~\ref{fig:fig3} we present $\mathcal{E}\left(\Delta /d\right)$ as dependent on interplanar distance for various $n$ and for $x=0.05$. It is remarkable that all the curves are symmetric with respect to the centre of the film, i.e. $\mathcal{E}\left(\omega\right)=\mathcal{E}\left(n-\omega\right)$. It is a result of the similar symmetry of the RKKY coupling integral in ultrathin film (note that the symmetry properties of a RKKY coupling for a thin film have been mentioned first in the work of Wojtczak\cite{WojtczakRKKY}). Let us remember that the summation over coordination zones within the atomic plane can be performed using the two sets of coordination radii and coordination numbers different for odd and even values of $\omega$. However, for low values of charge carrier concentration the inverse of the Fermi wavevector is quite large and we are in the quasi-continuous regime so that the detailed lattice structure in the atomic plane is not of great importance. Thus, the symmetry $\mathcal{E}\left(\omega\right)=\mathcal{E}\left(n-\omega\right)$ is exact for every value of $\rho$ only for $n$ even, while it is approximate for low $\rho$ for $n$ odd. 

In view of the experimental data, it might be instructive to mention that the constant $C/a^2$  approximately amounts to 1.6 mJ/m$^{2}$ for (Ga,Mn)As. 

It is visible that for $n=5$ to 9, the interplanar couplings are ferromagnetic (F) for the whole range of distances (with an exclusion of wery weak antiferromagnetic (AF) values for $n=9$, near the centre of the film). When $n$ increases from 5 to 9, the coupling values are reduced continuously but the shape of the curve remains essentially unmodified. When $n$ is changed from 9 to 10, a QSE is visible (compare Fig.~\ref{fig:fig2}). Thus, the curve shape changes discontinuously with two minimas, separated by a central maximum. The coupling values are at the same time shifted towards stronger ferromagnetic values. Then, further increase of $n$ causes the gradual lowering of couplings and for $n=16$ the two minima become AF in character. They reach the maximum depth for $n=20$. Switching to $n=21$, the next manifestation of QSE is visible as the curve develops an additional central minimum between two maximas. Once more, the coupling values became shifted towards F, and the two AF minimas become shallower. When increasing $n$ up to $n=32$, we notice the same tendency as before, namely the couplings are lowered, the central minimum becomes AF in character and the AF couplings near the two other minimas tend to rise. 

It is instructive to compare the values of interplanar energy obtained by performing the summation in the formula (\ref{eq:interlayer2}) for the thin-film RKKY coupling integral (given by (\ref{eq:rkky_tf})) with the some summation for the ordinary 3D RKKY coupling (with the same nonzero width of contact potential).\cite{RKKYdiffused} The results are presented in Fig.~\ref{fig:fig4} for $\rho=x/n$, $n=20$ and $x=0.05$. The full circles correspond to the results derived from thin-film RKKY coupling, while the empty circles depict the interplanar coupling for 3D RKKY integral. It is visible that for the distances smaller that a half of film thickness, the 3D coupling values are shifted towards F coupling, and also some phase shift is present. The curve for 3D RKKY does not possess the symmetry with respect to the centre of the film so that it vanishes quite fast for $z/d>0.5$, unlike the proper thin-film RKKY coupling. In the inset of Fig.~\ref{fig:fig4}, the values of the exchange integral itself for $r=0$ for ultrathin film (solid line) and bulk 3D (dashed line) are plotted as a function of the distance $z$ between the interacting spins. Here one can easily observe the phase shift and the difference in magnitudes between the two curves.

The interlayer coupling energy for $n_{L}=1$, calculated on the basis of formulas (\ref{eq:interlayer},\ref{eq:interlayer2}) is a key point in our study. Knowing the interplanar coupling energies as a function of concentration of carriers and geometry of the ultrathin film, it is possible to search for the sets of parameters which result in the most robust antiferromagnetic interlayer coupling. As can be seen from the formula (\ref{eq:interlayer2}), the interlayer energy is calculated by summing the interplanar coupling energies for $2n_{L}-1$ interplanar distances, ranging from $\left(n_{S}+1\right)d$ up to $\left(2n_{L}+n_{S}-1\right)d$, with appropriate weights. Therefore, if the maximum number of $\mathcal{E}\left(z\right)$ values, starting from $z=\left(n_{S}+1\right)d$ is antiferromagnetic, the strongest antiferromagnetism can be expected. On the other hand, since the coupling for $z=\left(n_{S}+1\right)d$ enters the sum in (\ref{eq:interlayer2}) with the largest weight $n_{L}$, it is also advisable to select $n_{S}$ in order to have this AF coupling considerably strong. For example, from the analysis of the Fig.~\ref{fig:fig3}, prepared for $x=0.05$, we can expect to obtain the AF interplanar coupling for the first minimum when we select $n_{S}=3$ for $n=20$. For thicker films, $n=32$, the optimum value seems to be $n_{S}=4$ (the first minimum) and $n_{S}=14$ (minimum in the middle, noticing that for $n_{S}=13$ the most important coupling $\mathcal{E}\left(\left(n_{S}+1\right)d\right)$ would be only slightly antiferromagnetic). For the larger number of subsequent antiferromagnetic values for $\mathcal{E}\left(z\right)$, we expect the stability of AF interlayer coupling for thicker magnetic layers. This thickness is of importance for two reasons. For one thing, with a diluted system in mind, the larger $n_{L}$ means larger value of total magnetization; for another, maximizing the critical temperature is also advantageous. On the other hand, it appears that the increase of the concentration of charge carriers in the system would cause the antiferromagnetic behaviour to diminish. This is due to the fact that the distance between the subsequent zeros of the function $\mathcal{E}\left(z\right)$ decreases when $\rho$ increases, so that the function oscillates more rapidly and the chance of having AF couplings for a few subsequent interplanar distances $z$ decreases. Let us note that the last observation seems to stand in apparent contradiction to the general idea that for small concentration of charge carriers ferromagnetism is favoured (because the first zero of the exchange integral is shifted towards larger distances and more coordination zones belong to this ferromagnetic range). 

This effect is illustrated in the Fig.~\ref{fig:fig5}, where the interplanar couplings are plotted for three charge carrier concentrations: $\rho=x/n$, $\rho=4x/n$ and $\rho=8x/n$, for $n=20$ and $x=0.05$. It might be useful to mention here that these values for (Ga,Mn)As would correspond to the (physically relevant) hole concentrations of 1.1, 4.4 and 8.9$\cdot10^{20}$ cm$^{-3}$, respectively. It is clearly visible that for the lowest concentration of charge carriers, the antiferromagnetic minima are quite remarkable. When $\rho$ is increased fourfold or eightfold, the oscillations of coupling versus interplanar distance become much faster, so that only two or one coupling is of distinct AF character. 

\begin{figure}
\includegraphics[scale=0.71]{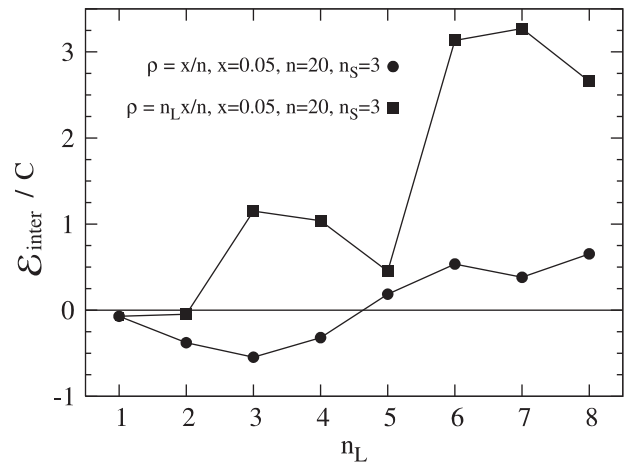}
\caption{Interlayer coupling energy for the system characterized by $x=0.05$, $n=20$ and $n_{S}=3$, as a function of magnetic layer thickness. The two charge carriers concentrations: $\rho=x/n$ (circles) and $\rho_{L}x/n$ (squares) are assumed.}
\label{fig:fig6}
\end{figure}

\begin{figure}
\includegraphics[scale=0.71]{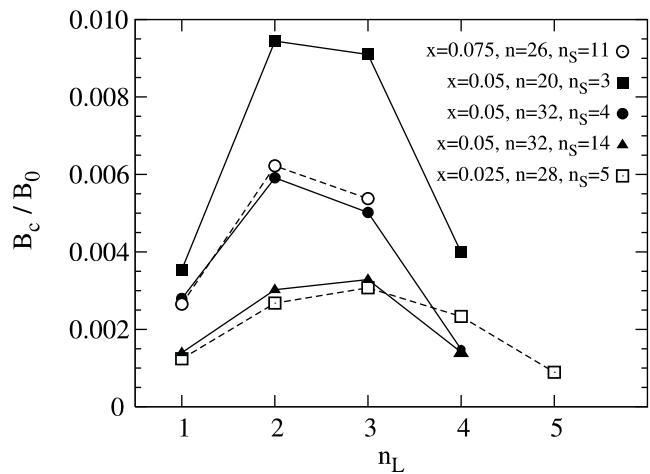}
\caption{External magnetic field required to switch the magnetizations of the layers from AF to F state as a function of layer thickness. The results are presented for $T\to 0$ and selected parameters, where AF interlayer coupling is predicted.}
\label{fig:fig7}
\end{figure}

Fig.~\ref{fig:fig6} illustrates the magnetic layer thickness dependence of the interlayer coupling energy for the already mentioned case of $x=0.05$, $n=20$ and $n_{S}=3$. The coupling energy is plotted for two different carrier concentration regimes. The circles correspond to the constant carrier concentration of $\rho=x/n$, independent on the thickness of the magnetic layers $n_{L}$. The second result, marked by squares, is for the carrier concentration $\rho_{L}x/n$ proportional to the number of magnetically doped atomic planes. It can be seen that for the second case, the initial antiferromagnetic interlayer coupling energy is lost soon when $n_{L}$ increases. Although the changes are strongly non-monotonic, yet the AF coupling is never restored. The situation is very different for the constant charge carrier concentration. The AF coupling persists up to $n_{L}=4$ and reaches the maximum strength for $n_{L}=3$. Further increase in magnetic layer thickness also leads to change to F interlayer coupling. The similar behaviour can be observed for another sets of parameters regarded as beneficial in terms of assuring persistent AF interlayer coupling.

Fig.~\ref{fig:fig7}  presents the critical external field $B_{\rm c}=\mu_{0}H_{\rm c}$ required to switch the direction of layer magnetizations from antiferromagnetic to ferromagnetic with temperature $T\to 0$. The values have been calculated from the formula (\ref{eq:Bcr}) for the parameters predicting the strongest antiferromagnetic interlayer coupling. In order to provide some reference point, we calculated the value of a normalization constant $B_{0}$ (\ref{eq:B0}) for a representative DMS, (Ga,Mn)As, which equals $B_{0}\simeq76$ T, so that the critical fields would lie in the range of hundreds of milliteslas. It is visible that the critical field for each case considered is largest for the magnetic layer thickness $n_{L}$=2 or 3. If the lowest value of critical field is needed, together with the highest N\'{e}el temperature, then the best choice seems to be the maximum thickness of the magnetic layer which still provides AF interlayer coupling. 

In connection with the predictions of Fig.~\ref{fig:fig7}, we would like to mention that an attempt to swith the magnetization direction in antiferromagnetically-coupled multilayer of GaAs/(Ga,Mn)As has been made by Chung \emph{et al}.\cite{chungAF} In their experiment, the field necessary to reverse the magnetization has been estimated at about 10 mT (judging from the magnetization curve and neutron reflectivity data); however, the full ferromagnetic alignment has not been achieved until 100 mT. It is worth noticing that in the experiment of Chung \emph{et al}. the spacer thickness was 12 ML, and each magnetic layer was 25 ML thick. 
\begin{figure}
\includegraphics[scale=0.71]{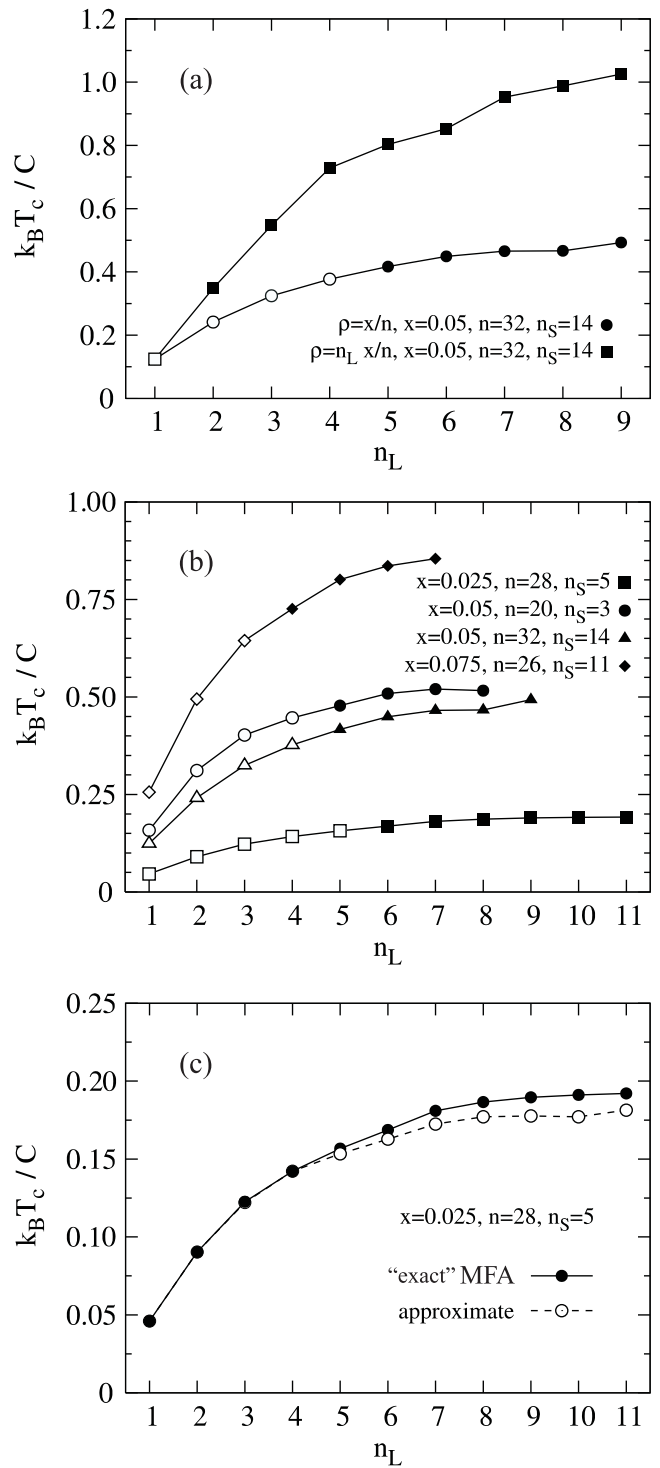}
\caption{Critical temperatures in MFA approximation for an ultrathin film with two magnetic layers. In (a) and (b) the filled symbols denote the Curie temperatures, while the empty ones are for the N\'{e}el temperatures. (a) Comparison of the results for charge carrier concentration $\rho_{L}x/n$ (squares) and $\rho=x/n$ (circles). (b) Results for $\rho=x/n$ and various parameter ranges which are advantageous for AF interlayer coupling. (c) Comparison of the results obtained from the approximate Eq.(15) (empty symbols) and without this approximation (filled symbols).}
\label{fig:fig8}
\end{figure}
\begin{figure}
\includegraphics[scale=0.85]{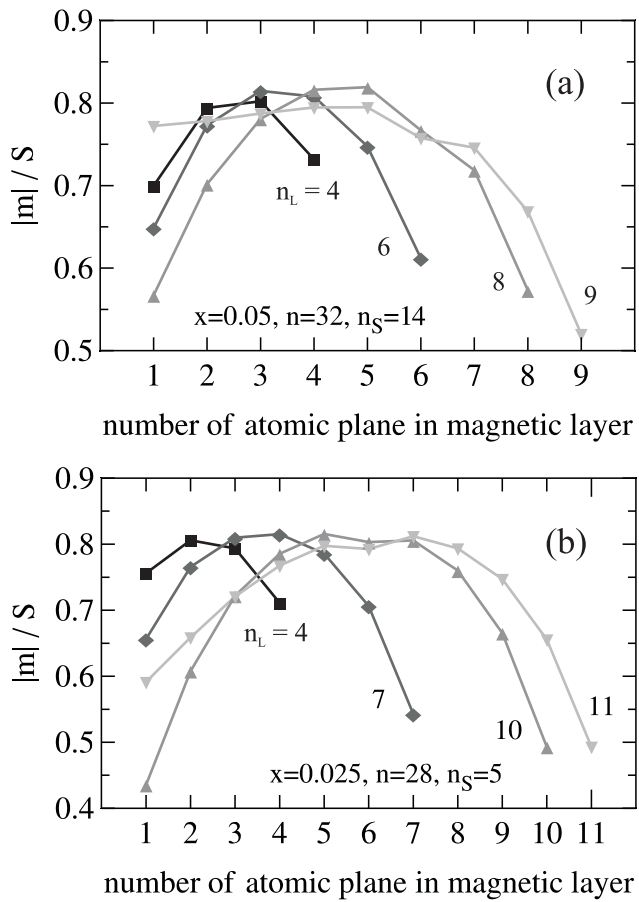}
\caption{Spatial distribution of magnetizations in subsequent monolayers, for various thicknesses $n_L$ and for two sets of parameters: (a) $x=0.05$, $n=32$ and $n_{S}=14$, and (b) $x=0.025$, $n=28$ and $n_{S}=5$. The temperature is $T=\left(2/3\right)T_{\rm c}$.}
\label{fig:fig9}
\end{figure}

In the studies of critical temperature of the system, it was necessary to accept a specific spin value. Therefore, we selected $S=5/2$, which is relevant for DMS. The behaviour of critical temperature of the system for the parameters leading to AF interlayer coupling is presented in Fig.~\ref{fig:fig8}(a). There, we plot the values of the critical temperatures for $x=0.05$, $n=32$ and $n_{S}=14$. Like in the Fig.~\ref{fig:fig5}, the circles correspond to constant charge carriers density $\rho=x/n$, while the squares are for $\rho_{L}x/n$. The empty symbols indicate the N\'{e}el temperatures (what corresponds to AF interlayer coupling), while the full symbols are for Curie temperature (F coupling). It is visible that for fixed $\rho$, the critical temperature tends to saturate when increasing $n_{L}$ and reaches approximately constant value for the magnetic layer consisting of several ML. It is seen that the N\'{e}el temperature for $n_{L}=4$ is very close to that saturated value. For $\rho_{L}x/n$, the critical temperatures are mostly Curie temperatures and rise relatively fast when the magnetic layer thickness increases.

In order to investigate more carefully the critical temperatures for the regime of fixed charge carriers concentration $\rho=x/n$, we perform the calculations for some other sets of parameters advantageous to AF coupling. The results are presented in Fig.~\ref{fig:fig8}(b). The same tendency for the critical temperature to saturate for thick magnetic layers is observed, and the maximum values of N\'{e}el temperature lie close to that limit of saturation. 

Let us mention that a similar behaviour of Curie temperature for thin films with RKKY interaction has been found in the Monte-Carlo based studies of Boselli \emph{et al.},\cite{boselli1} for fixed charge carriers concentration. However, they considered a thin film with magnetic impurities distributed randomly in all the atomic planes and their study was focused mainly on non-collinear magnetic phases, resembling canted ferromagnetism.

Fig.~\ref{fig:fig8}(c) compares of the critical temperature calculated from the "exact" MFA formula (\ref{eq:tc}) and from the approximate formula (\ref{eq:tcapproximate}), as dependent on magnetic layer thickness, for the set of parameters $x=0.025$, $n=28$ and $n_{S}=5$. It can be concluded that the assumption of uniform magnetization distribution inside each magnetic layer does not lead to a noticeable change in the critical temperature for the layers up to 4 ML thick. However, for thicker layers, the critical temperature is underestimated by an approximate formula in the order of 5 \%, which conclusion holds also for other sets of parameters.

It follows from the numerical calculations that the intralayer energy is much larger than the absolute value of the interlayer coupling. As a consequence, it is clearly visible from Eq.(\ref{eq:tcapproximate}) that the main contribution to the critical temperature originates from $\mathcal{E}_{\rm intra}$.

As mentioned previously, in the ordered state, each atomic plane of the magnetic layer has its own magnetization value and some non-uniform magnetization distribution exists across the layer thickness. To illustrate this behaviour, we calculated such  magnetization profiles for various thicknesses of the magnetic layers $n_{L}$ for the cases of $x=0.05$, $n=32$ and $n_{S}=14$ (Fig.~\ref{fig:fig9}(a)) as well as for $x=0.025$, $n=28$ and $n_{S}=5$ (Fig.~\ref{fig:fig9}(b)). The plots present a magnetization distribution in a single magnetic layer. Each profile is calculated for the temperature of $\frac{2}{3}T_{\rm c}$ for the given value of $n_{L}$, versus the number of the atomic plane counted starting from the plane situated closest to the undoped spacer. For the second magnetic layer, the values of $m$ are either the same (for F interlayer coupling) or they are of opposite sign (for AF coupling), so that the profile is either symmetric or antisymmetric with respect to the centre of the undoped spacer. It is noticeable that for small thicknesses, $n_{L}=2$, the distribution is almost uniform, while the increasing thickness of the magnetic layer causes the profile to reshape. The maximum of magnetization occurs for the atomic plane in the middle of the magnetic layer, while the smallest values are reached for the planes which are the most distant or the closest to the nonmagnetic spacer. For $n_{L}=8$ in Fig.~\ref{fig:fig9}(a) or $n_{L}=10$ Fig.~\ref{fig:fig9}(b), the distribution is approximately symmetric with respect to the middle atomic plane in the layer. It is noticeable that the differences in magnetizations for various atomic planes are strongly evident, since the magnetization at the boundaries of the magnetic layer is only a half of the maximum value reached in its centre. Interestingly, the profile losts its symmetry when we reach the maximum of possible magnetic layer thickness,  $n_{L}=9$ for (a) and $n_{L}=11$ for (b). Then the magnetization value in the atomic planes in the vicinity of the spacer is enhanced in comparison with the layers at the opposite side of the magnetic layer. Such an effect is more striking for the set of parameters $x=0.05$, $n=32$, $n_{S}=14$, where the distribution of magnetization switches from almost symmetric shape with respect to the middle plane of the layer (for $n_{L}=8$) to almost flat one with some drop at high distances from the spacer (for $n_{L}=9$). By approaching the critical temperature the profiles flatten and their magnitude is vanishing.

\section{Conclusion}

The analysis of a model magnetic bilayer with ultrathin film-modified RKKY interaction revealed an importance of QSE for obtaining antiferromagnetic interlayer coupling. Some ranges of parameters, advantageous for AF were identified, with a general remark that the carrier density should not be so high to make the oscillation period excessively short. The coupling is possible even for relatively thick spacers (14 ML for example); however, the maximum thickness of each magnetic layer should not exceed a few ML. The critical temperatures were found to saturate  when increasing magnetic layer thickness.  

The general conclusion which can be drawn from the study is that the occurrence of AF coupling presents a quite unique situation in comparison with the most common F ordering. Thus, in order to observe the AF configuration the special set of parameters has to be chosen including the impurity concentration, its distribution within the thin film, the thickness and the carriers density. Studies of sensitivity of the obtained configuration to the external magnetic field might be useful for practical realization of the switching function between the AF and F states. On the other hand, taking into consideration the importance of selecting the spacer thickness, some analysis embracing the interface roughness might also be interesting.

Although the magnetization profiles in thin films were already studied for NN interactions a long time ago\cite{Wojtczakprofile,Wiatrowski,Staniucha}, as far as we know for the long-range RKKY interactions the above presented profiles are found for the first time. Prediction of the profiles would need some experimental confirmation in the future studies of DMS systems. Let us observe that the presence of non-uniform magnetization profile causes a noticeable decrease in the total magnetization of each magnetic layer at finite temperatures (with respect to the prediction of the usual, bulk Brillouin function). 

In an ultrathin film, the characteristic features of RKKY exchange integral depend crucially on the standing-wave form of the free-carrier wavefunctions. They originate from the interference of waves propagating perpendicularly to the thin film plane. Thus, long enough phase coherence length for free charge carriers is required for this picture to be valid. Taking (Ga,Mn)As as an example, the coherence lengths of the order of 100 nm have been reported in quantum wires and rings for millikelvin temperatures (in presence of magnetic impurities),\cite{wagner} even though a mean free path for the carriers is shorter, down to an order of a lattice constant in the metallic-like regime.~\cite{sadowski} Such a value of phase coherence length exceeds the thicknesses of the films presented in this paper.

In the present work it has been assumed that the (random) positions of the magnetic impurity ions are uncorrelated. However, it could be also possible to investigate the influence of correlated positional disorder on the magnetic properties of the thin-film systems by considering the non-zero Warren-Cowley parameter. 

\begin{acknowledgments}
We gratefully acknowledge computational support on HUGO cluster at Department of Theoretical Physics and Astrophysics, P. J. \v{S}af\'{a}rik University in Ko\v{s}ice (with the help of M. Ja\v{s}\v{c}ur and Z. Kuscsik).
\end{acknowledgments}


\begin{thebibliography}{45}
\expandafter\ifx\csname natexlab\endcsname\relax\def\natexlab#1{#1}\fi
\expandafter\ifx\csname bibnamefont\endcsname\relax
  \def\bibnamefont#1{#1}\fi
\expandafter\ifx\csname bibfnamefont\endcsname\relax
  \def\bibfnamefont#1{#1}\fi
\expandafter\ifx\csname citenamefont\endcsname\relax
  \def\citenamefont#1{#1}\fi
\expandafter\ifx\csname url\endcsname\relax
  \def\url#1{\texttt{#1}}\fi
\expandafter\ifx\csname urlprefix\endcsname\relax\def\urlprefix{URL }\fi
\providecommand{\bibinfo}[2]{#2}
\providecommand{\eprint}[2][]{\url{#2}}

\bibitem[{\citenamefont{Gr\"unberg et~al.}(1986)\citenamefont{Gr\"unberg,
  Schreiber, Pang, Brodsky, and Sowers}}]{Grunberg}
\bibinfo{author}{\bibfnamefont{P.}~\bibnamefont{Gr\"unberg}},
  \bibinfo{author}{\bibfnamefont{R.}~\bibnamefont{Schreiber}},
  \bibinfo{author}{\bibfnamefont{Y.}~\bibnamefont{Pang}},
  \bibinfo{author}{\bibfnamefont{M.~B.} \bibnamefont{Brodsky}},
  \bibnamefont{and} \bibinfo{author}{\bibfnamefont{H.}~\bibnamefont{Sowers}},
  \bibinfo{journal}{Phys. Rev. Lett.} \textbf{\bibinfo{volume}{57}},
  \bibinfo{pages}{2442} (\bibinfo{year}{1986}).

\bibitem[{\citenamefont{Gr{\"{u}}nberg}(2001)}]{grunbergJMMM}
\bibinfo{author}{\bibfnamefont{P.}~\bibnamefont{Gr{\"{u}}nberg}},
  \bibinfo{journal}{J. Magn. Magn. Mat.} \textbf{\bibinfo{volume}{226-230}},
  \bibinfo{pages}{1688} (\bibinfo{year}{2001}).

\bibitem[{\citenamefont{Bruno}(1999)}]{BrunoC}
\bibinfo{author}{\bibfnamefont{P.}~\bibnamefont{Bruno}},
  \bibinfo{journal}{J. Phys.: Condens. Matter}
  \textbf{\bibinfo{volume}{11}}, \bibinfo{pages}{9403} (\bibinfo{year}{1999}); \bibinfo{journal}{Phys. Rev. B} \textbf{\bibinfo{volume}{52}},
  \bibinfo{pages}{411} (\bibinfo{year}{1995}).

\bibitem[{\citenamefont{Barna\ifmmode~\acute{s}\else
  \'{s}\fi{}}(1995)}]{Barnas1}
\bibinfo{author}{\bibfnamefont{J.}~\bibnamefont{Barna\'{s}}}, \bibinfo{journal}{Phys. Rev. B} \textbf{\bibinfo{volume}{52}},
  \bibinfo{pages}{10744} (\bibinfo{year}{1995}); \bibinfo{journal}{Phys. Rev. B} \textbf{\bibinfo{volume}{54}},
  \bibinfo{pages}{12332} (\bibinfo{year}{1996}).

\bibitem[{\citenamefont{Sapozhnikov and Genkin}(1995)}]{Sapozhnikov}
\bibinfo{author}{\bibfnamefont{M.}~\bibnamefont{Sapozhnikov}} \bibnamefont{and}
  \bibinfo{author}{\bibfnamefont{G.}~\bibnamefont{Genkin}},
  \bibinfo{journal}{J. Magn. Magn. Mat.}
  \textbf{\bibinfo{volume}{139}}, \bibinfo{pages}{179} (\bibinfo{year}{1995}).

\bibitem[{\citenamefont{Dugaev et~al.}(1999)\citenamefont{Dugaev, Litvinov,
  Dobrowolski, and Story}}]{dugaev}
\bibinfo{author}{\bibfnamefont{V.}~\bibnamefont{Dugaev}},
  \bibinfo{author}{\bibfnamefont{V.}~\bibnamefont{Litvinov}},
  \bibinfo{author}{\bibfnamefont{W.}~\bibnamefont{Dobrowolski}},
  \bibnamefont{and} \bibinfo{author}{\bibfnamefont{T.}~\bibnamefont{Story}},
  \bibinfo{journal}{Solid State Commun.} \textbf{\bibinfo{volume}{110}},
  \bibinfo{pages}{351} (\bibinfo{year}{1999}).

\bibitem[{\citenamefont{Edwards et~al.}(1991)\citenamefont{Edwards, Mathon,
  Muniz, and Phan}}]{Edwards2}
\bibinfo{author}{\bibfnamefont{D.~M.} \bibnamefont{Edwards}},
  \bibinfo{author}{\bibfnamefont{J.}~\bibnamefont{Mathon}},
  \bibinfo{author}{\bibfnamefont{R.~B.} \bibnamefont{Muniz}}, \bibnamefont{and}
  \bibinfo{author}{\bibfnamefont{M.~S.} \bibnamefont{Phan}},
  \bibinfo{journal}{Phys. Rev. Lett.} \textbf{\bibinfo{volume}{67}},
  \bibinfo{pages}{493} (\bibinfo{year}{1991}).

\bibitem[{\citenamefont{Staniucha and Urbaniak-Kucharczyk}(2006)}]{Staniucha}
\bibinfo{author}{\bibfnamefont{I.}~\bibnamefont{Staniucha}} \bibnamefont{and}
  \bibinfo{author}{\bibfnamefont{A.}~\bibnamefont{Urbaniak-Kucharczyk}},
  \bibinfo{journal}{Phys. Status Solidi C} \textbf{\bibinfo{volume}{3}},
  \bibinfo{pages}{65} (\bibinfo{year}{2006}).

\bibitem[{\citenamefont{J.Fabian et~al.}(2007)\citenamefont{J.Fabian,
  Matos-Abiague, Ertler, Stano, and \v{Z}uti\'{c}}}]{Fabian}
\bibinfo{author}{\bibnamefont{J.}~\bibnamefont{Fabian}},
  \bibinfo{author}{\bibfnamefont{A.}~\bibnamefont{Matos-Abiague}},
  \bibinfo{author}{\bibfnamefont{C.}~\bibnamefont{Ertler}},
  \bibinfo{author}{\bibfnamefont{P.}~\bibnamefont{Stano}}, \bibnamefont{and}
  \bibinfo{author}{\bibfnamefont{I.}~\bibnamefont{\v{Z}uti\'{c}}},
  \bibinfo{journal}{Acta Phys. Slovaca} \textbf{\bibinfo{volume}{57}},
  \bibinfo{pages}{565} (\bibinfo{year}{2007}).

\bibitem[{\citenamefont{Fert}(2008)}]{Fert}
\bibinfo{author}{\bibfnamefont{A.}~\bibnamefont{Fert}}, \bibinfo{journal}{Thin
  Solid Films} \textbf{\bibinfo{volume}{517}}, \bibinfo{pages}{2}
  (\bibinfo{year}{2008}).

\bibitem[{\citenamefont{T.Dietl et~al.}(1997)\citenamefont{T.Dietl, Haury, and
  d'Aubign\'{e}}}]{DietlPRB}
\bibinfo{author}{\bibnamefont{T.}~\bibnamefont{Dietl}},
  \bibinfo{author}{\bibfnamefont{A.}~\bibnamefont{Haury}}, \bibnamefont{and}
  \bibinfo{author}{\bibfnamefont{Y.~M.} \bibnamefont{d'Aubign\'{e}}},
  \bibinfo{journal}{Phys. Rev. B} \textbf{\bibinfo{volume}{55}},
  \bibinfo{pages}{R3347} (\bibinfo{year}{1997}); \bibinfo{author}{\bibnamefont{T.Dietl}},
  \bibinfo{author}{\bibfnamefont{J.}~\bibnamefont{Cibert}},
  \bibinfo{author}{\bibfnamefont{D.}~\bibnamefont{Ferrand}}, \bibnamefont{and}
  \bibinfo{author}{\bibfnamefont{Y.~M.} \bibnamefont{d'Aubign\'{e}}},
  \bibinfo{journal}{Mater. Sci. Eng.}
  \textbf{\bibinfo{volume}{B 63}}, \bibinfo{pages}{103} (\bibinfo{year}{1999}).

\bibitem[{\citenamefont{Giddings et~al.}(2006)\citenamefont{Giddings,
  Jungwirth, and Gallagher}}]{giddingspssc}
\bibinfo{author}{\bibfnamefont{A.~D.} \bibnamefont{Giddings}},
  \bibinfo{author}{\bibfnamefont{T.}~\bibnamefont{Jungwirth}},
  \bibnamefont{and} \bibinfo{author}{\bibfnamefont{B.~L.}
  \bibnamefont{Gallagher}}, \bibinfo{journal}{Phys. Status Solidi C}
  \textbf{\bibinfo{volume}{3}}, \bibinfo{pages}{4070} (\bibinfo{year}{2006}); \bibinfo{author}{\bibfnamefont{A.~D.} \bibnamefont{Giddings}},
  \bibinfo{author}{\bibfnamefont{T.}~\bibnamefont{Jungwirth}},
  \bibnamefont{and} \bibinfo{author}{\bibfnamefont{B.~L.}
  \bibnamefont{Gallagher}}, \bibinfo{journal}{Phys. Rev. B}
  \textbf{\bibinfo{volume}{78}}, \bibinfo{eid}{165312}
  (\bibinfo{year}{2008}).

\bibitem[{\citenamefont{Sankowski and Kacman}(2005)}]{Sankowskicoupling}
\bibinfo{author}{\bibfnamefont{P.}~\bibnamefont{Sankowski}} \bibnamefont{and}
  \bibinfo{author}{\bibfnamefont{P.}~\bibnamefont{Kacman}},
  \bibinfo{journal}{Phys. Rev. B} \textbf{\bibinfo{volume}{71}},
  \bibinfo{eid}{201303(R)} (\bibinfo{year}{2005}).

\bibitem[{\citenamefont{Jungwirth et~al.}(1999)\citenamefont{Jungwirth,
  Atkinson, Lee, and MacDonald}}]{Jungwirthcoupling}
\bibinfo{author}{\bibfnamefont{T.}~\bibnamefont{Jungwirth}},
  \bibinfo{author}{\bibfnamefont{W.~A.} \bibnamefont{Atkinson}},
  \bibinfo{author}{\bibfnamefont{B.~H.} \bibnamefont{Lee}}, \bibnamefont{and}
  \bibinfo{author}{\bibfnamefont{A.~H.} \bibnamefont{MacDonald}},
  \bibinfo{journal}{Phys. Rev. B} \textbf{\bibinfo{volume}{59}},
  \bibinfo{pages}{9818} (\bibinfo{year}{1999}).

\bibitem[{\citenamefont{Akiba et~al.}(1998)\citenamefont{Akiba, Matsukura,
  Shen, Ohno, Ohno, Oiwa, Katsumoto, and Iye}}]{akibacoupling}
\bibinfo{author}{\bibfnamefont{N.}~\bibnamefont{Akiba}},
  \bibinfo{author}{\bibfnamefont{F.}~\bibnamefont{Matsukura}},
  \bibinfo{author}{\bibfnamefont{A.}~\bibnamefont{Shen}},
  \bibinfo{author}{\bibfnamefont{Y.}~\bibnamefont{Ohno}},
  \bibinfo{author}{\bibfnamefont{H.}~\bibnamefont{Ohno}},
  \bibinfo{author}{\bibfnamefont{A.}~\bibnamefont{Oiwa}},
  \bibinfo{author}{\bibfnamefont{S.}~\bibnamefont{Katsumoto}},
  \bibnamefont{and} \bibinfo{author}{\bibfnamefont{Y.}~\bibnamefont{Iye}},
  \bibinfo{journal}{Appl. Phys. Lett.} \textbf{\bibinfo{volume}{73}},
  \bibinfo{pages}{2122} (\bibinfo{year}{1998}).

\bibitem[{\citenamefont{Mathieu et~al.}(2002)\citenamefont{Mathieu, Svedlindh,
  Sadowski, Swiatek, Karlsteen, Kanski, and Ilver}}]{mathieucoupling}
\bibinfo{author}{\bibfnamefont{R.}~\bibnamefont{Mathieu}},
  \bibinfo{author}{\bibfnamefont{P.}~\bibnamefont{Svedlindh}},
  \bibinfo{author}{\bibfnamefont{J.}~\bibnamefont{Sadowski}},
  \bibinfo{author}{\bibfnamefont{K.}~\bibnamefont{Swiatek}},
  \bibinfo{author}{\bibfnamefont{M.}~\bibnamefont{Karlsteen}},
  \bibinfo{author}{\bibfnamefont{J.}~\bibnamefont{Kanski}}, \bibnamefont{and}
  \bibinfo{author}{\bibfnamefont{L.}~\bibnamefont{Ilver}},
  \bibinfo{journal}{Appl. Phys. Lett.} \textbf{\bibinfo{volume}{81}},
  \bibinfo{pages}{3013} (\bibinfo{year}{2002}).

\bibitem[{\citenamefont{Szuszkiewicz et~al.}(2003)\citenamefont{Szuszkiewicz,
  Dynowska, F., Hennion, Jouanne, J.F, and Sadowski}}]{szuszkiewiczcoupling1}
\bibinfo{author}{\bibfnamefont{W.}~\bibnamefont{Szuszkiewicz}},
  \bibinfo{author}{\bibfnamefont{E.}~\bibnamefont{Dynowska}},
  \bibinfo{author}{\bibfnamefont{F.}~\bibnamefont{Ott}},
  \bibinfo{author}{\bibfnamefont{B.}~\bibnamefont{Hennion}},
  \bibinfo{author}{\bibfnamefont{M.}~\bibnamefont{Jouanne}},
  \bibinfo{author}{\bibfnamefont{J.~F.}~\bibnamefont{Morhange}}, \bibnamefont{and}
  \bibinfo{author}{\bibfnamefont{J.}~\bibnamefont{Sadowski}},
  \bibinfo{journal}{J. Supercond.} \textbf{\bibinfo{volume}{16}}, \bibinfo{pages}{209}
  (\bibinfo{year}{2003}); \bibinfo{author}{\bibfnamefont{W.}~\bibnamefont{Szuszkiewicz}},
  \bibinfo{author}{\bibfnamefont{E.}~\bibnamefont{Dynowska}},
  \bibinfo{author}{\bibfnamefont{F.}~\bibnamefont{Ott}},
  \bibinfo{author}{\bibfnamefont{B.}~\bibnamefont{Hennion}},
  \bibinfo{author}{\bibfnamefont{M.}~\bibnamefont{Jouanne}},
  \bibinfo{author}{\bibfnamefont{J.~F.}~\bibnamefont{Morhange}},
  \bibinfo{author}{\bibfnamefont{M.}~\bibnamefont{Karlsteen}},
  \bibnamefont{and} \bibinfo{author}{\bibfnamefont{J.}~\bibnamefont{Sadowski}},
  \bibinfo{journal}{Acta Phys. Pol. A} \textbf{\bibinfo{volume}{100}},
  \bibinfo{pages}{335} (\bibinfo{year}{2001}).


\bibitem[{\citenamefont{Kirby et~al.}(2007)\citenamefont{Kirby, Borchers, Liu,
  Ge, Cho, Dobrowolska, and Furdyna}}]{kirbycoupling}
\bibinfo{author}{\bibfnamefont{B.~J.} \bibnamefont{Kirby}},
  \bibinfo{author}{\bibfnamefont{J.~A.} \bibnamefont{Borchers}},
  \bibinfo{author}{\bibfnamefont{X.}~\bibnamefont{Liu}},
  \bibinfo{author}{\bibfnamefont{Z.}~\bibnamefont{Ge}},
  \bibinfo{author}{\bibfnamefont{Y.~J.} \bibnamefont{Cho}},
  \bibinfo{author}{\bibfnamefont{M.}~\bibnamefont{Dobrowolska}},
  \bibnamefont{and} \bibinfo{author}{\bibfnamefont{J.~K.}
  \bibnamefont{Furdyna}}, \bibinfo{journal}{Phys. Rev. B}
  \textbf{\bibinfo{volume}{76}}, \bibinfo{eid}{205316}
  (\bibinfo{year}{2007}).

\bibitem[{\citenamefont{Ge et~al.}(2007)\citenamefont{Ge, Zhou, Cho, Liu,
  Furdyna, and Dobrowolska}}]{AFcouplingjunctions}
\bibinfo{author}{\bibfnamefont{Z.}~\bibnamefont{Ge}},
  \bibinfo{author}{\bibfnamefont{Y.~Y.} \bibnamefont{Zhou}},
  \bibinfo{author}{\bibfnamefont{Y.-J.} \bibnamefont{Cho}},
  \bibinfo{author}{\bibfnamefont{X.}~\bibnamefont{Liu}},
  \bibinfo{author}{\bibfnamefont{J.~K.} \bibnamefont{Furdyna}},
  \bibnamefont{and}
  \bibinfo{author}{\bibfnamefont{M.}~\bibnamefont{Dobrowolska}},
  \bibinfo{journal}{Appl. Phys. Lett.} \textbf{\bibinfo{volume}{91}},
  \bibinfo{eid}{152109} (\bibinfo{year}{2007}).

\bibitem[{\citenamefont{Chung et~al.}(2008)\citenamefont{Chung, Chung, Lee,
  Kirby, Borchers, Cho, Liu, and Furdyna}}]{chungAF}
\bibinfo{author}{\bibfnamefont{J.-H.} \bibnamefont{Chung}},
  \bibinfo{author}{\bibfnamefont{S.~J.} \bibnamefont{Chung}},
  \bibinfo{author}{\bibfnamefont{S.}~\bibnamefont{Lee}},
  \bibinfo{author}{\bibfnamefont{B.~J.} \bibnamefont{Kirby}},
  \bibinfo{author}{\bibfnamefont{J.~A.} \bibnamefont{Borchers}},
  \bibinfo{author}{\bibfnamefont{Y.~J.} \bibnamefont{Cho}},
  \bibinfo{author}{\bibfnamefont{X.}~\bibnamefont{Liu}}, \bibnamefont{and}
  \bibinfo{author}{\bibfnamefont{J.~K.} \bibnamefont{Furdyna}},
  \bibinfo{journal}{Phys. Rev. Lett.} \textbf{\bibinfo{volume}{101}},
  \bibinfo{eid}{237202} (\bibinfo{year}{2008}).

\bibitem[{\citenamefont{K{\"{o}}nig et~al.}(2002)\citenamefont{K{\"{o}}nig,
  Schliemann, Jungwirth, and MacDonald}}]{Schliemann}
\bibinfo{author}{\bibfnamefont{J.}~\bibnamefont{K{\"{o}}nig}},
  \bibinfo{author}{\bibfnamefont{J.}~\bibnamefont{Schliemann}},
  \bibinfo{author}{\bibfnamefont{T.}~\bibnamefont{Jungwirth}},
  \bibnamefont{and}
  \bibinfo{author}{\bibfnamefont{A.}~\bibnamefont{MacDonald}}, in
  \emph{\bibinfo{booktitle}{Electronic Structure and Magnetism of Complex
  Materials}}, edited by \bibinfo{editor}{\bibnamefont{D.J.Singh}}
  \bibnamefont{and} \bibinfo{editor}{\bibnamefont{D.A.Papaconstantopoulos}}
  (\bibinfo{publisher}{Springer}, \bibinfo{year}{2002}).

\bibitem[{\citenamefont{Schliemann et~al.}(2001)\citenamefont{Schliemann,
  K\"onig, and MacDonald}}]{SchliemannMC}
\bibinfo{author}{\bibfnamefont{J.}~\bibnamefont{Schliemann}},
  \bibinfo{author}{\bibfnamefont{J.}~\bibnamefont{K\"onig}}, \bibnamefont{and}
  \bibinfo{author}{\bibfnamefont{A.~H.} \bibnamefont{MacDonald}},
  \bibinfo{journal}{Phys. Rev. B} \textbf{\bibinfo{volume}{64}},
  \bibinfo{pages}{165201} (\bibinfo{year}{2001}); \bibinfo{author}{\bibfnamefont{J.}~\bibnamefont{Schliemann}} \bibnamefont{and}
  \bibinfo{author}{\bibfnamefont{A.~H.} \bibnamefont{MacDonald}},
  \bibinfo{journal}{Phys. Rev. Lett.} \textbf{\bibinfo{volume}{88}},
  \bibinfo{pages}{137201} (\bibinfo{year}{2002}); \bibinfo{author}{\bibfnamefont{J.}~\bibnamefont{Schliemann}} \bibnamefont{and}
  \bibinfo{author}{\bibfnamefont{A.~H.} \bibnamefont{MacDonald}},
  \bibinfo{journal}{J. Supercond.} \textbf{\bibinfo{volume}{16}}, \bibinfo{pages}{11}
  (\bibinfo{year}{2003}); \bibinfo{author}{\bibfnamefont{J.}~\bibnamefont{Schliemann}},
  \bibinfo{journal}{Phys. Rev. B} \textbf{\bibinfo{volume}{67}},
  \bibinfo{pages}{045202} (\bibinfo{year}{2003}).

\bibitem[{\citenamefont{Brey and G\'omez-Santos}(2003)}]{Brey}
\bibinfo{author}{\bibfnamefont{L.}~\bibnamefont{Brey}} \bibnamefont{and}
  \bibinfo{author}{\bibfnamefont{G.}~\bibnamefont{G\'omez-Santos}},
  \bibinfo{journal}{Phys. Rev. B} \textbf{\bibinfo{volume}{68}},
  \bibinfo{pages}{115206} (\bibinfo{year}{2003}).

\bibitem[{\citenamefont{Sza\l{}owski and
  Balcerzak}(2008{\natexlab{a}})}]{RKKYdiffused}
\bibinfo{author}{\bibfnamefont{K.}~\bibnamefont{Sza\l{}owski}}
  \bibnamefont{and}
  \bibinfo{author}{\bibfnamefont{T.}~\bibnamefont{Balcerzak}},
  \bibinfo{journal}{Phys. Rev. B} \textbf{\bibinfo{volume}{78}},
  \bibinfo{eid}{024419} (\bibinfo{year}{2008}{\natexlab{a}}).

\bibitem[{\citenamefont{Sza\l{}owski and
  Balcerzak}(2008{\natexlab{b}})}]{Jaszowiec}
\bibinfo{author}{\bibfnamefont{K.}~\bibnamefont{Sza\l{}owski}}
  \bibnamefont{and}
  \bibinfo{author}{\bibfnamefont{T.}~\bibnamefont{Balcerzak}},
  \bibinfo{journal}{Acta Phys. Pol. A} \textbf{\bibinfo{volume}{114}},
  \bibinfo{pages}{1375} (\bibinfo{year}{2008}{\natexlab{b}}).

\bibitem[{\citenamefont{Balcerzak}(2006{\natexlab{a}})}]{BalcerzakTF2}
\bibinfo{author}{\bibfnamefont{T.}~\bibnamefont{Balcerzak}},
  \bibinfo{journal}{Thin Solid Films} \textbf{\bibinfo{volume}{515}},
  \bibinfo{pages}{2814} (\bibinfo{year}{2006}{\natexlab{a}}).

\bibitem[{\citenamefont{Balcerzak}(2007{\natexlab{a}})}]{BalcerzakJMMM}
\bibinfo{author}{\bibfnamefont{T.}~\bibnamefont{Balcerzak}},
  \bibinfo{journal}{J. Magn. Magn. Mat.} \textbf{\bibinfo{volume}{310}},
  \bibinfo{pages}{1651} (\bibinfo{year}{2007}{\natexlab{a}}).

\bibitem[{\citenamefont{Balcerzak}(2007{\natexlab{b}})}]{BalcerzakRKKY}
\bibinfo{author}{\bibfnamefont{T.}~\bibnamefont{Balcerzak}}, in
  \emph{\bibinfo{booktitle}{Trends in thin solid films research}}, edited by
  \bibinfo{editor}{\bibfnamefont{A.~R.} \bibnamefont{Jost}}
  (\bibinfo{publisher}{Nova Science Publishers},
  \bibinfo{year}{2007}{\natexlab{b}}), chap.~\bibinfo{chapter}{9}.

\bibitem[{\citenamefont{Balcerzak}(2006{\natexlab{b}})}]{BalcerzakTF1}
\bibinfo{author}{\bibfnamefont{T.}~\bibnamefont{Balcerzak}},
  \bibinfo{journal}{Thin Solid Films} \textbf{\bibinfo{volume}{500}},
  \bibinfo{pages}{341} (\bibinfo{year}{2006}{\natexlab{b}}).

\bibitem[{\citenamefont{Edwards and Loveluck}(1970)}]{edwards}
\bibinfo{author}{\bibfnamefont{S.~F.} \bibnamefont{Edwards}} \bibnamefont{and}
  \bibinfo{author}{\bibfnamefont{J.~M.} \bibnamefont{Loveluck}},
  \bibinfo{journal}{J. Phys. C: Metal Physics Suppl.}
  \textbf{\bibinfo{volume}{3}}, \bibinfo{pages}{S261} (\bibinfo{year}{1970}).

\bibitem[{\citenamefont{Sza\l{}owski and
  Balcerzak}(2008{\natexlab{c}})}]{RKKYcorrelations}
\bibinfo{author}{\bibfnamefont{K.}~\bibnamefont{Sza\l{}owski}}
  \bibnamefont{and}
  \bibinfo{author}{\bibfnamefont{T.}~\bibnamefont{Balcerzak}},
  \bibinfo{journal}{Phys. Rev. B} \textbf{\bibinfo{volume}{77}},
  \bibinfo{eid}{115204} (\bibinfo{year}{2008}{\natexlab{c}}).

\bibitem[{\citenamefont{Wojtczak}(1969)}]{WojtczakRKKY}
\bibinfo{author}{\bibfnamefont{L.}~\bibnamefont{Wojtczak}},
  \bibinfo{journal}{Acta Phys. Pol.} \textbf{\bibinfo{volume}{36}},
  \bibinfo{pages}{585} (\bibinfo{year}{1969}).

\bibitem[{\citenamefont{Boselli et~al.}(2003)\citenamefont{Boselli,
  da~Cunha~Lima, and Ghazali}}]{boselli1}
\bibinfo{author}{\bibfnamefont{M.~A.} \bibnamefont{Boselli}},
  \bibinfo{author}{\bibfnamefont{I.~C.} \bibnamefont{da~Cunha~Lima}},
  \bibnamefont{and} \bibinfo{author}{\bibfnamefont{A.}~\bibnamefont{Ghazali}},
  \bibinfo{journal}{Phys. Rev. B} \textbf{\bibinfo{volume}{68}},
  \bibinfo{pages}{085319} (\bibinfo{year}{2003}).

\bibitem[{\citenamefont{Wojtczak}(1970)}]{Wojtczakprofile}
\bibinfo{author}{\bibfnamefont{L.}~\bibnamefont{Wojtczak}},
  \bibinfo{journal}{Czech. J. Phys.} \textbf{\bibinfo{volume}{B20}},
  \bibinfo{pages}{247} (\bibinfo{year}{1970}).

\bibitem[{\citenamefont{Wiatrowski et~al.}(1986)\citenamefont{Wiatrowski,
  Balcerzak, Wojtczak, and Mielnicki}}]{Wiatrowski}
\bibinfo{author}{\bibfnamefont{G.}~\bibnamefont{Wiatrowski}},
  \bibinfo{author}{\bibfnamefont{T.}~\bibnamefont{Balcerzak}},
  \bibinfo{author}{\bibfnamefont{L.}~\bibnamefont{Wojtczak}}, \bibnamefont{and}
  \bibinfo{author}{\bibfnamefont{J.}~\bibnamefont{Mielnicki}},
  \bibinfo{journal}{Phys. Status Solidi B} \textbf{\bibinfo{volume}{138}},
  \bibinfo{pages}{189} (\bibinfo{year}{1986}).

\bibitem[{\citenamefont{Wagner et~al.}(2006)\citenamefont{Wagner, Neumaier,
  Reinwald, Wegscheider, and Weiss}}]{wagner}
\bibinfo{author}{\bibfnamefont{K.}~\bibnamefont{Wagner}},
  \bibinfo{author}{\bibfnamefont{D.}~\bibnamefont{Neumaier}},
  \bibinfo{author}{\bibfnamefont{M.}~\bibnamefont{Reinwald}},
  \bibinfo{author}{\bibfnamefont{W.}~\bibnamefont{Wegscheider}},
  \bibnamefont{and} \bibinfo{author}{\bibfnamefont{D.}~\bibnamefont{Weiss}},
  \bibinfo{journal}{Phys. Rev. Lett.} \textbf{\bibinfo{volume}{97}},
  \bibinfo{pages}{056803} (\bibinfo{year}{2006}); \bibinfo{author}{\bibfnamefont{D.}~\bibnamefont{Neumaier}},
  \bibinfo{author}{\bibfnamefont{K.}~\bibnamefont{Wagner}},
  \bibinfo{author}{\bibfnamefont{U.}~\bibnamefont{Wurstbauer}},
  \bibinfo{author}{\bibfnamefont{M.}~\bibnamefont{Reinwald}},
  \bibinfo{author}{\bibfnamefont{W.}~\bibnamefont{Wegscheider}},
  \bibnamefont{and} \bibinfo{author}{\bibfnamefont{D.}~\bibnamefont{Weiss}},
  \bibinfo{journal}{New J. Phys.} \textbf{\bibinfo{volume}{10}},
  \bibinfo{pages}{055016} (\bibinfo{year}{2008}).

\bibitem[{\citenamefont{S{\o}rensen et~al.}(2003)\citenamefont{S{\o}rensen,
  Lindelof, Sadowski, Mathieu, and Svedlindh}}]{sadowski}
\bibinfo{author}{\bibfnamefont{B.~S.} \bibnamefont{S{\o}rensen}},
  \bibinfo{author}{\bibfnamefont{P.~E.} \bibnamefont{Lindelof}},
  \bibinfo{author}{\bibfnamefont{J.}~\bibnamefont{Sadowski}},
  \bibinfo{author}{\bibfnamefont{R.}~\bibnamefont{Mathieu}}, \bibnamefont{and}
  \bibinfo{author}{\bibfnamefont{P.}~\bibnamefont{Svedlindh}},
  \bibinfo{journal}{Appl. Phys. Lett.} \textbf{\bibinfo{volume}{82}},
  \bibinfo{pages}{2287} (\bibinfo{year}{2003}).

\end{thebibliography}

\end{document}